\documentstyle[preprint,eqsecnum,aps]{revtex}

\begin{document}
\draft
\preprint{}
\title{Spontaneous Flux and Magnetic Interference Patterns
in 0-$\pi$ Josephson Junctions}
\author{J.R. Kirtley}
\address{ IBM T.J. Watson Research Center, P.O. Box 218,
Yorktown Heights, NY 10598}
\author{K.A. Moler}
\address{ Department of Physics, Princeton University, Princeton,
NJ 08544}
\author{D.J. Scalapino}
\address{Department of Physics, University of California,
Santa Barbara, CA 93106}
\date{\today}
\maketitle
\begin{abstract}
The spontaneous flux generation and magnetic field modulation of the
critical current in a $0 - \pi$ Josephson junction are calculated for
different ratios of the junction length $L$ to the Josephsen penetration
depth $\lambda_J$, and different ratios of the $0$-junction length to the
$\pi$-junction length. These calculations apply to
a Pb-YBCO c-axis oriented junction with one YBCO twin
boundary, as well as other experimental systems. Measurements of such
a junction
can provide information on the nature of the $c$-axis Josephson
coupling and the symmetry of the order parameter in YBCO.
We find spontaneous flux even for very
short symmetric $0-\pi$ junctions, but
asymmetric junctions have qualitatively different behavior.
\end{abstract}
\pacs{73.40.Gk,73.40.Rw,74.50.+r,74.72.-h}

\narrowtext

\section{Introduction}

Although many measurements now support the idea that the gap in the
high-temperature cuprate superconductors has dominant
$d_{x^{2}-y^{2}}$ character\cite{drefs},
an orthorhombic superconductor such as
$YBa_2Cu_3O_{7 - \delta}$ (YBCO) cannot be described as a purely
tetragonal $d$-wave. In the orthorhombic symmetry group the
$d_{x^{2}-y{2}}-$ and $s-$wave basis function belong to the
same irreducible representation, so that one expects that the gap
can be described as $d_{x^{2}-y^{2}}-$wave with some $s-$wave admixture,
$d+ \alpha s$. The deviation from purely $d$-wave
behavior is supported by the observations\cite{sun,kleiner}
of Josephson pair tunneling from Pb into c-axis oriented YBCO, although
the variation of the Josephson tunneling strength with twin density,
the relative areas of the twins, and the possible role of Josephson
screening currents, remain open questions\cite{carbotte}. It has also
been proposed that these results can be explained by the development of
a complex order parameter at twin boundaries
\cite{sigrist}, or mixing of states with $s$ and $d_{x^{2}-y^{2}}$
symmetry at the Pb-YBCO interface\cite{bahcall}.
Part of the problem of interpreting these experiments is
the complex geometry presented by highly twinned materials.
Here we consider
the simpler case of a c-axis YBCO-Pb tunnel junction in which
the YBCO has
only one twin boundary in the junction. Experimental efforts
on this system are currently underway\cite{clarke}.

In the idealized geometry we describe below (Fig. 1), the Pb/YBCO
junction
which contains a single twin boundary can be described as a ``$0-\pi$''
junction, where the pair transfer integral has a relative pair phase of
$0$ in the left part of the junction (length $L_{0}$) and $\pi$ in the
right part of the junction (length $L_{\pi}$).
Here we are interested in understanding the behavior of such ``$0- \pi$''
Josephson junctions.
While we describe our geometry in terms
of c-axis tunneling in YBCO-Pb junctions, our results apply equally
well to
the $0-\pi$ a-b plane YBCO-Pb planar junctions\cite{wollman2} and
grain boundary junctions\cite{blnkt,miller}
used for phase sensitive tests of the symmetry
of the high-$T_c$ superconductors. In particular, while traditionally one
considers the magnetic interference pattern, we will also be interested
in spontaneous flux,
flux generated by the self-screening
Josephson currents in the absence of a drive current and an external
field
$H_{e}$, which can be directly probed by, for example,
a scanning SQUID microscope\cite{blnkt}.
As noted in Ref. \cite{xu}, the two measurements are complementary,
since while the magnetic
field interference patterns are only strikingly different between a
$0-\pi$ junction
and a conventional
$0-0$ junction in the short junction limit
$L \stackrel{<}{\sim} \lambda_J$,
appreciable spontaneous
flux only appears in the $0-\pi$ junction in the long junction limit
$L \stackrel{>}{\sim}  \lambda_{J}$.

The maximum Josephson current $I_{c}$ which a junction can carry
versus an external magnetic field $H_{e}$ applied parallel to the
plane of the
junction is called the ``magnetic interference pattern''. The
interference pattern
for a short ($L \ll \lambda_J$)
$0 - \pi$ junction
has been discussed by Wollman {\it et al.}\cite{wollman2}.
Xu {\it et al.}\cite{xu} have estimated the crossover to the
limiting case of a long
($L \gg \lambda_J$) $0 -\pi$ junction. With the approximations made by Xu
{\it et al.}, the magnetically modulated critical current of a
long $0 - \pi$ junction
is basically identical to the conventional $0-0$ junction due
to entrance of
a half-flux quantum vortex, and the solution
which contains spontaneously generated flux ceases to be a global
minimum of the free energy for short junctions. Our exact numerical
solutions,
however, show that there is still a ``dip'' in the center of the
diffraction patterns
even for junctions as long as $10 \lambda_{J}$, and that the
symmetric junction
should contain spontaneously generated magnetic flux even for
$L < \lambda_{J}$.
Here we are primarily
interested in determining the behavior of a $0-\pi$ junction for
intermediate values of $L/ \lambda_J$  and different $L_{0}/L_{\pi}$
asymmetries. We believe that the dependence on $L/\lambda_J$ of
both the field-modulated
critical current and the spontaneous flux generation of a Pb-YBCO
junction with one twin boundary
can provide important information on the $c$-axis coupling.

In the following, we discuss a numerical method for calculating the
spontaneous flux generation and magnetic interference patterns
for arbitrary
junction lengths $L/ \lambda_J$ and junction asymmetries $L_{0}/L_{\pi}$.
We then compare results for the critical current $I_c$ versus
external magnetic
field $H_{e}$ for the case of a traditional $0-0$ junction with
the symmetric
$0-\pi$ case as a function of the reduced junction length $L/\lambda_J$.
Following this, we focus on the Josephson screening currents
by examining the flux generated by these self-currents in the absence of
an external current and magnetic field. As previously noted,
this self generated
flux can be probed
using a scanning SQUID microscope\cite{blnkt}, providing
important information on
the nature of the c-axis Josephson coupling. We then turn
to the interesting
case in which the areas of the two twin regions differ and
again look at the
self-flux versus  $L/\lambda_J$. There is a qualitative
difference between the
results for the spontaneous flux generation which occurs in
asymmetric junctions
compared to symmetric
junctions. We show briefly how these results can be extended
to a junction with
multiple twin boundaries.
We conclude by discussing what this type of experiment can tell us
about the Pb-YBCO c-axis Josephson tunneling.

\section{The Calculation}

The geometry and gap structure which we envision is illustrated in Fig. 1.
In this idealized geometry\cite{owen}, the upper YBCO strip contains a
twin boundary which separates the YBCO-Pb junction into two regions.
As schematically illustrated in Fig. 1, we have taken the phase of the
larger lobe of the $d+ \alpha s$ gap in region $0 < x < L_0$ to be
positive. Then, if the phase is locked across the grain boundary as one
expects\cite{walker},
the larger lobe in region $L_{0} < x < L$ will be negative. Thus
the pair transfer integral between the YBCO and the Pb leads
to a relative
Josephson pair phase of 0 for $0 < x < L_{0}$
and $\pi$ for $L_{0} < x < L$.

Following the notation of Owen and Scalapino\cite{owen}, we
describe a junction (Fig. 1) with width w small compared to $\lambda_J$
in the $y$-direction, of length
$L$ in the $x$-direction, carrying a total current $I_{tot}$
in the $x$-direction,
in an external magnetic field
$H_{e}$ oriented in the $y$-direction.
We include the effects of the $0$ and $\pi$ junctions by introducing
an extra phase angle $\theta(x)$ which takes the value $0$ or $\pi$.
The superconducting phase difference across the junction is then just
the solution of the Sine-Gordon equation:
\begin{equation}
\frac{\partial^{2}\phi}{\partial x^{2}} =
\frac{1}{\lambda_{J}^{2}}\sin{( \phi(x)+\theta(x) )}.
\end{equation}
Applying Ampere's Law with a contour of integration around the perimeter
of the junction in the $xy$ plane leads to the boundary condition
\begin{equation}
H(L)-H(0) = 4 \pi I_{tot}/c{\rm w}.
\end{equation}
Contours of integration in the $yz$ plane circling the leads at $x=0$ and
$x=L$ lead to
\begin{equation}
H(L)+H(0) = 2H_{e}.
\end{equation}
The Josephson penetration depth is given by
\begin{equation}
\lambda_J = \left (\frac{\hbar c^{2}}{8 \pi e d j_c} \right )^{1/2},
\end{equation}
where $d$ is the sum of the
Pb and YBCO $\lambda_{ab}$ penetration depths plus the thickness of the
insulator layer between the two superconductors,
and $j_{c}$ is the Josephson critical current density.
The current per unit length through the junction is given by:
\begin{equation}
j(x) = {\rm{w}} j_{c}sin(\phi(x)+\theta(x)).
\end{equation}
The gradient of the phase is in turn related to the field $H$ in the
junction by:
\begin{equation}
H(x) = \frac{\hbar c}{2ed} \frac{\partial \phi}{\partial x}.
\end{equation}
If we redefine the parameters $H(x)$ and  $I_{tot}$ as
\begin{equation}
h(x) = \frac{2ed\lambda_{J}}{\hbar c}H(x) = \lambda_{J} \frac{\partial
\phi}{\partial x},
\end{equation}
and
\begin{equation}
i_{tot} = \frac{I_{tot}}{2\lambda_{J} {\rm{w}} j_{c}},
\end{equation}
the boundary conditions become
\begin{equation}
h(L) = i_{tot}+h_{e}
\end{equation}
\begin{equation}
h(0) = -i_{tot}+h_{e}.
\end{equation}
Defining a dimensionless coupling parameter
\begin{equation}
\alpha = \frac{\lambda_{J}^{2}}{(\Delta x)^{2}},
\end{equation}
the differential equation turns into a difference equation on a grid
of size $\Delta x$:
\begin{equation}
{\phi_{n+1}-2\phi_{n}+\phi_{n-1}}
= \frac{1}{\alpha}\sin{(\phi_{n}+\theta_{n})}.
\end{equation}
The boundary conditions are then described as difference equations,
where $n_{j}$
is the total number of junctions:
\begin{equation}
\phi_{n_{j}}-\phi_{n_{j}-1}  = \frac{i_{tot}+h_{e}}{\sqrt{\alpha}}
\end{equation}
\begin{equation}
\phi_{2}-\phi_{1}  = \frac{-i_{tot}+h_{e}}{\sqrt{\alpha}}
\end{equation}
These coupled difference equations are solved using a relaxation
method to find the solution $\phi(x)$.
The free energy of this solution is given by:
\begin{equation}
F = \frac{\hbar j_{c} {\rm{w}}}{2e} \int_{0}^{L}\left( 1 -
\cos{(\phi(x)+\theta(x))}+
 \frac{\lambda_{J}^{2}}{2} \left( \frac{\partial \phi}{\partial x}
 \right)^{2} \right) dx
\end{equation}
Written as a difference equation, the free energy for the vortex
solution becomes:
\begin{equation}
F_{V} = \frac{\hbar j_{c} {\rm{w}} \Delta x}{2e}\sum_{1}^{n_{j}-1}
\left( 1-cos(\phi_{n}+\theta_{n})+\frac{\alpha}{2}
(\phi_{n+1}-\phi_{n})^{2} \right)
\end{equation}
while the free energy of the no-vortex ($\phi=0$ everywhere) solution
is:
\begin{equation}
F_{0} = \frac{\hbar j_{c} {\rm{w}} \Delta x}{2e}
\sum_{1}^{n_{j}-1} (1-\cos{\theta_{n}})
\end{equation}
To obtain the critical current, the current $I_{tot}$
through the junction
is increased in steps, iterating the solution until
either the convergence
criterion

\begin{displaymath}
\epsilon >  \{  [ ( \phi_{1} - \phi_{2} - ( i - h_{e})/\alpha^{1/2})^2 +
(\phi_{n_{j}} - \phi_{n_{j-1}} - ( i + h_{e} )/\alpha^{1/2})^2
\end{displaymath}
\begin{equation}
+ \sum_{n=2}^{n_{j}-1} (\phi_{n+1} +
\phi_{n-1} -2 \phi_{n} - sin( \phi_n + \theta_n)/
\alpha )^2  ] / n_{j}  \} ^ {1/2} / \pi
\end {equation}
is satisfied,
with $\epsilon$ normally chosen to be $10^{-6}$,
or until the solution diverges with $\mid \phi(n_{j}/2) \mid > 25$.
For particular
values of the parameters, we compare solutions with $\epsilon$ taken to
be $10^{-6}$, $10^{-7}$, and $10^{-8}$ to check the convergence
of the solution.
The largest value of $I_{tot}$ for which the iterations
converge is taken to be the critical current $I_c$. For the
numerical calculations
shown in the next section, we took $L=10$, $\Delta x = 0.1$,
$n_{j}=L/\Delta x
= 100$, which meant that the dimensionless coupling constant ranged from
$\alpha = 10^{4}$ for $L/\lambda_{J} = 1$ to
$\alpha = 10^{2}$ for $L/\lambda_{J}=10$: $\Delta x$ was always
much smaller than $L_{0}$, $L_{\pi}$, or $\lambda_{J}$.

\section{Results}

Fig. 2 compares our results for the magnetic interference
pattern for a $0-0$
junction (a) and a symmetric $0-\pi$ junction (b) for
various values of the
reduced length $L/\lambda_J$. Here we have
plotted $I_{c}/I_{1}$ versus $\Phi/\Phi_{0}$,
where $I_{1} = j_{c} \rm{w} L$,
$\Phi = H_{e}dL$, and $\Phi_{0}  = hc/2e$ is the
superconducting flux quantum.
In the short junction limit $L \ll \lambda_J$,
the critical current can be written as:
\begin{equation}
I_{c}(B) = \frac{I_{1}}{L} \max
\left (\int_{0}^{L} sin(\frac{2 \pi H_{e} x d}{\Phi_{0}} +
\theta(x) + \psi) dx \right),
\label{eq:pattern}
\end{equation}
where $\theta(x)$ can be $0$ or $\pi$, and the maximum value of
the expression is obtained by varying $0 < \psi < 2 \pi$
for each value of
$H_{e}$. For the $0-0$ junction
($\theta(x) = 0$) this expression reduces to:
\begin{equation}
\frac{I_{c}}{I_{1}} = \frac{\mid sin(\pi \Phi/ \Phi_0)
\mid}{\mid \pi \Phi / \Phi_0 \mid} ,
\end{equation}
while for the symmetric $0-\pi$ junction
($\theta(x) = 0, x < L/2; \theta(x) =
\pi, x \geq L/2$) this becomes:
\begin{equation}
\frac{I_{c}}{I_{1}} = \frac{sin^{2}(\pi
\Phi / 2 \Phi_{0})}{\mid \pi \Phi / 2 \Phi_0 \mid }.
\end{equation}
These equations are plotted as the solid lines in Figure 2.
Also included in Figure 2a
is the result of Owen and Scalapino for the reduced length
$L/ \lambda_{J} = 10$.
For the $0-0$ junction the effect of increasing $L$ is to
reduce the height of the
central peak in the magnetic interference pattern,
and to reduce the amplitude of
the successive oscillations as $H_{e}$ is increased.
For the $0- \pi$ junction, increasing $L$
tends to reduce the depth of the minimum at $H_{e}=0$,
and as $L/ \lambda_J
\rightarrow \infty $
the curves for the $0-0$ junction and the $0- \pi$
junction become identical as
discussed by Xu {\it et al.}\cite{xu}. However,
even for $L/ \lambda_{J} = 10$, we
find, as shown in Fig. 2(b), that the critical current
initially increases
with flux until $\Phi / \Phi_{0} \cong 0.5$. This differs from the
$L / \lambda_{J} = 10$ result shown in Ref. \cite{xu}.
We believe that this difference results from the approximation
made by Xu {\it et al.}
that the presence of a $\pi$ vortex simply changes the phase
in the $\pi$ section of
the junction by $\pi$, effectively changing it into a $0$-junction.
This approximation
is valid on length scales large compared to $\lambda_{J}$, but,
as shown in Fig. 2b,
leads to an experimentally detectable difference in the solutions
even for $L$ as
long as $10 \lambda_{J}$. For short and medium length junctions,
Xu {\it et al.}
took the analytical long-junction expression and made the
approximation that
the effect of shortening the junction would be simply to cut
off this expression.
However, this is not
a valid solution of the modified sine-Gordon equation and the boundary
conditions. It also leads, as we show below, to dramatic
differences in both
the free energy and the
total flux for the $\pi$-vortex solutions for short- to medium-length
junctions.

Fig. 3 addresses the question of spontaneous flux generation
in $0- \pi$ junctions
as a function of reduced length $L / \lambda_J$ and the degree
of asymmetry
($L_{\pi}/L$). Fig. 3a plots the ratio of the free energy of the
state with some
spontaneous flux to the state with no flux. The dashed line is
the result of
Xu {\it et al.}, using the approximation outlined above.
The solid lines are our full
results. Note that, in contrast to the results of Xu {\it et al.},
for the symmetric $0-\pi$ junction ($L_{\pi}/L = 0.5$) the
state with spontaneous flux always has lower energy than
the state with no flux,
and thus some self generated flux
should therefore be present for all values of $L / \lambda_J$.
Fig 3b plots the
spontaneous flux {\it vs.} $L_{\pi}/\lambda_{J}$.
The inset in this figure shows that the spontaneous flux
for a symmetric junction
follows a power law
dependence on junction length for short junctions. In this limit
the phase $\phi$ has only
small deviations from an average value of $\pi/2$, and the
spontaneously generated
field increases linearly towards the center of the junction
from a value of zero at
the edges.
A simple geometrical argument, expanding the Sine-Gordon equation about
$\phi = \pi/2$, implies that for short junctions the spontaneous
flux should be given by $\Phi/\Phi_{0} = (L/\lambda_{J})^{2}/8\pi$.
This relation is
plotted as a straight line in the inset of Fig. 3b. However, for
the asymmetric case, the state with no flux has the lowest free
energy for
short junctions, and as shown in Fig. 3b, there
is no spontaneously generated flux, up to a critical value of
$L_{\pi} / \lambda_J
\stackrel{<}{\sim} 1 $. The amount of spontaneously generated
flux approaches $\Phi_0 /2 $ as $L_{\pi}$ gets larger, and rate
of increase
of flux at the onset of flux generation increases
for less symmetrical junctions.

The onset of spontaneous flux generation is also apparent in the magnetic
interference patterns, as shown in Fig. 4. Here we hold the
asymmetry factor
$L_{\pi}/L$ fixed at 0.25, and vary the length of the junction
$L_{\pi}/ \lambda_{J}$.
For small $L_{\pi}/ \lambda_{J}$ there is no spontaneous flux
generated, and the
magnetic interference has a minimum at zero field. However, as
$L_{\pi} / \lambda_{J}$ approaches unity, there is an abrupt
shift to a magnetic
interference pattern with a maximum at zero field. The solid
line in Fig. 4 is
the short-junction limit calculated from Eq. 3.1,
which is a good indicator of the actual interference pattern
only until flux
generation becomes energetically favorable.

The behavior of junctions with multiple twin boundaries can
be calculated using the
same techniques, as long as the junction width w is small
compared to $\lambda_{J}$.
We show the results of one such set of calculations in Fig. 5.
Here we have assumed
that there are 100 twins in the junction, with the probability
of a particular twin
having $0-$ or $\pi-$phase randomly distributed with the probability of
$\pi$ being 0.25. The circles in
Fig. 5a show the randomly distributed set of $0-$ and $\pi$
phases used in this calculation. The solid lines in Fig. 5a
show the solution for
the phase angle $\phi$ at $H_{e}=0$ and $I=I_{c}$, for
various values of $L/\lambda_{J}$.
These curves show, as expected, that the gradient of $\phi$,
and therefore the flux
penetration into the junction, is spread throughout the
junction for small $L/\lambda_{J}$,
but is localised at the junction edges for $L/\lambda_{J} \gg 1$.
The solid line in Fig. 5b shows the predicted interference
pattern from Eq. 3.1 in the short-junction limit.
The other symbols show the result of
the full numerical solution for $L/ \lambda_{J} = 2,4,$ and $10$.
Even for this small
number of randomly distributed twins, the interference pattern,
although reduced in
critical current, is similar to that for a single $0$-junction (Fig. 2a)
especially for larger values of $L/ \lambda_{J}$. This figure shows
how important it is that complete knowledge of the distribution of
twins be available
for the correct interpretation of experimental work on $c$-axis tunneling
from twinned samples.

\section{Conclusion}

As discussed by Sun {\it et al.}\cite{sun}, the results of their Pb-YBCO
$c$-axis
Josephson tunneling experiments can be interpreted as showing
that the order
parameter in YBCO must have some $s$-wave component. While this could be
consistent with an order parameter having
$d_{x^{2}-y^{2}}+ \alpha s$ symmetry in an orthorhombic material, the
interpretation of the experiments done on twinned materials
depend on details of the multi-twinned geometry. Here we have
analyzed a simpler situation involving a single twin boundary.
In this case, if the
$d_{x^{2}-y^{2}}$ contribution is dominant, giving the type of order
parameters illustrated in Fig. 1, one should observe the behavior we have
discussed. In particular, if the junction areas for the two twin regions
are equal, one has the conventional minimum in the critical
current at zero
magnetic field as shown in Fig. 2(b), if $L/ \lambda_{J}$
is sufficiently small,
and provided the current density is uniform on a scale set
by $\lambda_{J}$. For the asymmetric case in which the areas of the
two twin regions differ, more striking behavior should appear.
There should be
a sudden shift of the interference pattern from one with a
minimum at zero field,
to one with a maximum (Fig. 4), as
$L_{\pi} / \lambda_{J}$ goes through unity.
One way to vary $L / \lambda_{J}$ would be by changing
$ \lambda_{J}$ with temperature. Another would be to apply a magnetic
field $H_{x}$ to vary $\lambda_{J}(H_{x}) =
[ \hbar c^{2}/(8 \pi e d I(H_{x})/{\rm w}L)]^{1/2}$,
with $I(H_{x}) = I_{1} \mid sin(\pi \Phi _{x}/\Phi_{0})
\mid / \mid \pi \Phi_{x}
/ \Phi_{0} \mid $ , and $\Phi_{x} = H_{x} d \rm{w}$.
Alternatively, or
in addition, one could use a scanning SQUID microscope to measure the
spontaneous flux generated by the self-screening Josephson currents.
For the
case of equal areas, this spontaneous flux rises continuously
as $L_{\pi} / \lambda_{J}$ increases. However, for nonequal
areas one finds no spontaneous flux generation below a critical value of
$L_{\pi} / \lambda_{J} \sim 1$ and then a sharp increase in flux
should occur.
We believe that
if these types of behavior are observed, they would show that the Pb-YBCO
$c$-axis Josephson coupling is consistent with a description
of the high $T_{c}$ superconductors as having dominant
$d_{x^{2}-y^{2}}$ symmetry.

\acknowledgments

We would like to thank J. Clarke, R.C. Dynes, and A.G. Sun for
interesting
discussions of their experiments prior to publication, and
S. Bahcall for a
careful reading of the manuscript.
DJS acknowledges the National Science
Foundation under grant No. DMR95-27304.

\begin{figure}
\caption{(a) Junction geometry showing the directions of the
external current flow
$I_{tot}$ and magnetic field $H_{e}$. The dashed line marks
the twin boundary, and the
gap functions are schematically illustrated such that the basic Josephson
pair phase arising from the pair wave function overlap is $0$ for
$0 < x < L_{0} $ and $\pi$ for $L_{0} < x < L$.
(b) Relative pair tunneling phase across
the junction as a function of position.}

\vspace{0.3in}
\caption{The
dependence of the junction critical current $I_{c}/I_{1}$ as
a function of the field in the junction  $\Phi / \Phi_{0}$,
for a series of
junction lengths $L / \lambda_J$, for a $0-0$ junction (a)
and a symmetric
$0- \pi$ junction (b). The solid curve is the analytical
result in the short junction
$L \ll \lambda_J$ limit. The dashed curve in (a) is the
result for $L/ \lambda_J = 10$
from Ref. [12]. Curves for successive values of $L/ \lambda_{J}$
have been offset
vertically by 0.5 units for clarity.}

\vspace{0.3in}
\caption{
The ratio of the free energy for the solution with spontaneous
flux $F_{v}$ to that with no flux $F_{0}$ (a), and the spontaneous
total flux in the junction $\Phi / \Phi_{0}$ (b), for a $0- \pi$ junction
as a function of the length
of the $\pi$ junction
$L_{\pi}/ \lambda_{J}$, for different asymmetries $L_{\pi}/L$.
The inset of (b) shows the results for a symmetric junction on
a log-log scale. The straight line is the relation $\Phi/\Phi_{0}
= (L/\lambda_{J})
^{2}/8\pi$.}

\vspace{0.3in}
\caption{The dependence of the critical current $I_{c}/I_{1}$ as a
function
of field in the junction $\Phi / \Phi_{0}$, for a junction with asymmetry
$L_{\pi}/L = 0.25$, for three different junction lengths:
$L_{\pi}/\lambda_{J} = 0.5$,
well below the critical length for spontaneous flux generation;
$L_{\pi}/\lambda_{J} =
0.75$, near the critical length; and $L_{\pi}/\lambda_{J}  = 1$,
above the critical length.
The solid curve is calculated using the short junction approximation,
Eq. (3.1). Curves for successive values of $L/ \lambda_{J}$
have been offset vertically by
0.5 units for clarity.}

\vspace{0.3in}
\caption{(a) The circles show
100 randomly distributed $0$ and $\pi$ phases, assuming a
probability of $\pi$-phase=0.25. The solid lines show the
solutions for $\phi$ at $I=I_{c}$
and $H_{e}=0$.
(b) The solid curve is the magnetic
interference pattern for a junction with this distribution of
phases in the
short junction limit, Eq. (3.1). The solid points are the
result of the full
numerical calculation for various values of $L/ \lambda_{J}$. Successive
$L/ \lambda_{J}$ curves have been offset vertically by 0.2
units for clarity.}

\label{autonum}
\end{figure}

\end{document}